\documentclass{article}
\usepackage{amssymb}

\newcommand{\beq}[1]{\begin{equation} \label{#1} }
\newcommand{\beqa}[1]{\begin{eqnarray} \label{#1} }
\newcommand{\eeq}{\end{equation}}
\newcommand{\eeqa}{\end{eqnarray}}
\newcommand{\ba}[1]{\begin{array}{#1}}
\newcommand{\ea}{\end{array}}
\newcommand{\rf}[1]{(\ref{#1})}
\newcommand{\bosy}[1]{ \mbox{\boldmath ${#1}$} }

\begin{document}

\title{ Thermoelastic Relaxation in Elastic Structures, \\ with Applications to Thin Plates}

\author{A. N. Norris$^{1}$ and D. M. Photiadis$^{2}$\\ \\   (1) Mechanical and Aerospace Engineering, 
	Rutgers University, \\ Piscataway NJ 08854-8058, USA norris@rutgers.edu \\ (2) Naval Research Laboratory, Code 7130, Washington, \\ DC 20375-5350, USA photiadis@nrl.navy.mil} 
	\date{}
\maketitle

\begin{abstract} 
A new result enables direct calculation of thermoelastic damping in  vibrating elastic 
solids.  The mechanism for energy loss is thermal diffusion caused by  inhomogeneous 
deformation, flexure in  thin plates.   The general result is combined  with the Kirchhoff 
assumption to obtain a new equation for the flexural vibration of thin  plates incorporating 
thermoelastic loss as a damping term.  The thermal relaxation loss is  inhomogeneous and 
depends upon the local state of vibrating flexure, specifically, the  principal curvatures at 
a given point on the plate.  Thermal loss is zero at points where the  principal curvatures 
are equal and opposite, that is, saddle shaped or pure anticlastic deformation.   
Conversely, 
loss is maximum at points where the curvatures are equal, that is, synclastic or  
spherical 
flexure.  The influence of modal curvature on the thermoelastic damping is  described through 
a modal participation factor.   The effect of transverse thermal  diffusion on plane 
wave propagation is also examined.  It is shown that transverse diffusion  effects are always 
small provided the plate thickness is far greater than the thermal phonon  mean free path, a 
requirement for the validity of the classical theory of heat transport. These  results 
generalize Zener's theory of thermoelastic loss in beams and are useful in  predicting mode 
widths in MEMS and NEMS oscillators.  
\end{abstract}

\section{Introduction}

High Q resonators are central to the development of new devices and  applications that 
include RF filters, next generation MRI systems, and torque magnetometers.    Silicon based 
micro- and nano- electromechanical systems (MEMS/NEMS) oscillators are the  candidates of 
choice, and include free standing planar devices, such as double paddle  oscillators (DPOs),  
and micro-cantilevers.  As the oscillators shrink in size, it has been  found that the Q 
achieved is orders of magnitude smaller than expected based on classical  fundamental loss 
mechanisms.  Many mechanisms have been proposed, including surface loss  
 \cite{carr,mohanty,yang02} that increases with the surface to volume ratio.   However, under 
 controlled conditions with minimal surface defects and adsorbates, measurements on silicon 
 DPOs have shown that room temperature losses are adequately described by thermoelastic 
 relaxation, while unexplained mechanisms operate at lower temperatures.  Interestingly, the 
 mode of vibration of DPOs is designed to be primarily torsional with very little flexure 
 (and hence no thermoelastic coupling).  However, as demonstrated by Photiadis and his 
 co-workers  \cite{Photiadis02,Liu02,Liu,Houston} it is precisely the small amount of 
flexural 
motion that accounts for loss in these supposedly torsional oscillators. 

 The purpose of this paper is to provide a consistent theory for  predicting intrinsic 
 dissipation arising from thermoelasticity in elastic structures.  Particular attention will 
 be given to flexural motion of thin plates.  This work is a step towards understanding the 
 fundamental limits of dissipation in small structures such as NEM and MEM devices.  

 We are concerned with determining the thermomechanical loss of elastic modes, for 
example, the 
 flexural mode of a rectangular thin plate.   A useful point of departure is the classic 
 theory of Zener \cite{Zener} for anelastic thermoelastic damping.   The key to this approach 
 is the assumption, implicit in Zener's work, that there is little relative  difference 
 between the isentropic (unrelaxed) and isothermal (relaxed) mechanical responses, and hence 
 the mechanical and thermal problems are essentially decoupled.  Since the thermoelasticity 
 is weak, the transition from the instantaneous or unrelaxed system to the relaxed state can 
 be viewed as a quasistatic thermal process, governed by the standard equations for thermal 
 diffusion, although now in the presence of an inhomogeneous deformation.  

 Energy loss in a mechanical oscillator is measured in terms of the quality factor,  defined 
 as $Q=2\pi {\cal E}_0/\Delta {\cal E}$, where ${\cal E}_0$ is the mechanical energy of the 
 oscillator and $\Delta {\cal E}$ is the energy loss per cycle. The quality factor for a 
 lightly damped single degree of freedom system with nondimensional damping $\zeta \ll 1$ is 
 $Q = 1/(2\zeta)$, and by assumption we only consider systems with light damping, or $Q \gg 1$.  
 The relation between $Q$ and $\alpha_{at}$,  the attenuation per unit length of a  
 propagating wave of frequency $\omega$,  is $Q = \omega /(2\alpha_{at} v) $, where $v$ is 
the 
 speed of energy propagation, also equivalent to the group velocity.  
 This identity can be derived by  assuming the energy is a quadratic function of the field 
 variables, so that energy decays with distance $d$ as $e^{-2\alpha_{at} d}$. The distance 
 travelled in one period is $d= 2\pi v/\omega$, and hence the fractional decrease in energy 
per 
 period is $1- e^{-4\pi \alpha_{at} v/\omega} \approx 4\pi \alpha_{at} v/\omega $ from which 
 the relation for $Q$ follows. 
  
 Thermoelastic loss can be most simply viewed as a  relaxation mechanism with a single 
 relaxation time $\tau$.  The generic  
 frequency dependent  quality factor $Q(\omega)$ for a relaxation mechanism is 
\beq{q1}
Q^{-1} = \frac{\Delta c}{c_0} \frac{\omega \tau}{1+ \omega^2 \tau^2}, 
\eeq
 where  $\Delta c$ is the (relatively) instantaneous increase in elastic modulus, $c_0 
 \rightarrow c_0 +\Delta c $, caused by the process under consideration.     The change in 
 elasticity is well known for thermoelasticity, and $\tau$ has been estimated for several 
configurations.  Thus, Zener \cite{Zener} showed  for flexure of a beam that 
\beq{q1a}
Q^{-1} = \frac{E\alpha^2 T}{C_p} \frac{\omega \tau_0}{1+ \omega^2 \tau_0^2},
\eeq
 where $E$ is the isothermal Young's modulus, $T$ the absolute temperature, $\alpha$ the 
 volume coefficient of thermal expansion, and $C_p$ is the heat capacity at constant stress.  
 The characteristic relaxation time is $\tau_0 = h^2C_p/(\pi^2 K)$ where $h$ is the thickness 
 and $K$ the thermal conductivity.  In fact, as Zener first demonstrated \cite{Zener37}, the 
 simple expression \rf{q1a} is the leading term in an infinite series which is well 
 approximated by the single term (see  \rf{mod} and Appendix B). 
 Zener's method was derived in the context of scalar problems, where the strain, for 
 instance, involves a single component.  An important example is the flexure of a beam or 
 reed, as considered originally by Zener, and later by others, for  example,~\cite{Nowick}.  
The  present work generalizes Zener's method to consider general elastic deformation.  This 
 includes the inhomogeneous deformation associated with modes in thin plates and other 
structures. 

 Since the original work of Zener numerous papers have appeared on thermal relaxation in the 
 context of coupled thermoelasticity.  In a pair of papers \cite{Alblas61,Alblas81} Alblas 
 provided a rigorous formulation using continuum thermomechanics and linear elasticity theory 
 for isotropic materials. He derived  detailed and explicit solutions for the thermoelastic 
 damping in vibrating beams,  including the circular rod and the rectangular beam.  The 
 result for the latter was derived separately by Lifshitz and Roukes \cite{Lifshitz00}, 
 although Alblas' solution is the more general of the two.  These analyses are compared with 
 the present formulation later (see Appendix B). 
 Kinra and Milligan \cite{Kinra94} again derived the coupled isotropic thermoelastic 
 equations and provided a solution for  unidimensional structures, including a discontinuous 
 beam.  Perhaps the most thorough analysis of thermal damping in the context of the coupled  
 equations of thermoelasticity is due to Chadwick \cite{Chadwick62}.  By considering a modal 
 decomposition of the elastic and thermal fields, an exact relation for the complex valued 
 frequency of oscillation of each mode was obtained.  This enabled Chadwick to derive a 
 generalization of Zener's expression for the thermoelastic damping of an arbitrary elastic 
 body.   Chadwick subsequently derived the governing equations of thermoelasticity for thin 
 plates and beams \cite{Chadwick62b}.  The equations are in the form of coupled equations, 
 one of which reduces to the classical equations for the structural mode, for example, 
flexural waves 
 in thin plates, and the other  the temperature diffusion equation, in the limit of zero 
 coupling.   The analysis in \cite{Chadwick62,Chadwick62b} is  restricted to isotropic 
solids.    

 This paper has several objectives.  The first is to demonstrate how the Zener model 
 follows from the full equations for the coupled dynamic system by using a consistent 
 approximation scheme.    In the process we  generalize Zener's approach to incorporate 
 general elastic deformation, specifically the elastic stress and strain tensors.  
 The main applications are to thin plate structures, for which we obtain a Zener-like result 
 for arbitrary flexural deformation that includes the general curvature tensor.  Our results 
 will also include the possibility of thermal diffusion in the lateral direction in thin 
 plates, which is explicitly ignored in Zener's approach, but was considered by Alblas 
 \cite{Alblas61,Alblas81} and Chadwick \cite{Chadwick62b}.  However, it will be shown that  
 circumstances under which lateral thermal flux becomes important coincide with the limit in 
 which the thin plate theory is no longer applicable.  

 The paper is arranged as follows. Governing equations of thermoelasticity are presented in 
 Section 2 for anisotropic elastic bodies.  General solutions are discussed in Section 3 with 
 no particular type of structure in mind.  The theory is applied to thin plates in flexure in 
 Section \ref{sec4} and  a non-dimensional  modal participation factor (MPF) is introduced 
 which defines the local contribution to thermoelastic (TE) loss in terms of the plate 
 curvature.  An  alternative method for deriving the TE loss of travelling flexural waves is 
 presented in Section \ref{sec5} using generalized  plate equations.  The effects of lateral 
 thermal diffusion are discussed in the context of travelling wave solutions in 
Section~\ref{sec5}.

\section{Equations of thermoelasticity}\label{sec2}

\subsection{Constitutive relations of thermomechanics}

The thermomechanical variables are the bulk stress and strain, 
 {\boldmath$\sigma$} and ${\bf e}$,    the temperature deviation, $\theta$ from the ambient 
 absolute temperature $\theta_a$ ($|\theta| /\theta_a \ll 1$), and the entropy deviation per 
 unit volume, $s$, from the ambient entropy $s_a$.  All quantities are
 defined relative to their ambient values, and would be zero in the absence of
exterior motivating forces.  The  constitutive
 relations for strain and entropy in terms of  the independent variables stress and 
temperature, $\{ \bosy{\sigma},\, \theta\}$,  are \cite{Christensen,Nowacki,Chand}
\beqa{2.01}
{\bf e} = {\bf S}  \mbox{\boldmath$\sigma$}  
        + \mbox{\boldmath$\alpha$}   \theta, ~~~~~~
s = C_p (\theta /\theta_a) + \mbox{\boldmath$\alpha$}\cdot   
\mbox{\boldmath$\sigma$}   .  
\eeqa
 Table~\ref{tableI} summarizes the alternative formulations of the same equations based on 
different 
 choices of the independent variables: $\{ {\bf e},\theta \}$, $\{ {\bf e},s\}$ or $\{ 
\bosy{\sigma}$, s\}.  
\begin{table}
%%%%%%%%%%%%%%%%%%%%    TABLE  I
\begin{center}
{\small
\begin{tabular}{|c|c|c|c|c|} \hline  %\hline
Dependent & $\bosy{\sigma}$,\,$\theta$&{\bf e},\,$\theta$&{\bf 
e},\,$s$&$\bosy{\sigma}$,\,$s$\\
variable  &        &    &          &      \\ \hline
  & & & & \\
${\bf e} = \mbox{}$& ${\bf S}\mbox{\boldmath$\sigma$}+\mbox{\boldmath$\alpha$}\theta$ 
    &    &   &   ${\bf S}_s \bosy{\sigma} +\frac{\theta_a}{C_p} \bosy{\alpha} \, s$ \\
  & & & &  \\ 
$s =\mbox{}$& $\frac{C_p}{\theta_a}\theta + \mbox{\boldmath$\alpha$}\cdot 
\mbox{\boldmath$\sigma$}$  &  $\frac{C_v}{\theta_a}\theta + { \mbox{\boldmath$\beta$} 
}\cdot {\bf e}$  &     &  \\
  & & & &   \\
 $\bosy{\sigma} =\mbox{}$ &    &  $ {\bf C} {\bf e} -  {  \mbox{\boldmath$\beta$} } \theta$   
& ${\bf C}_s {\bf e} -  \frac{\theta_a}{C_v}\mbox{\boldmath$\beta$} s$  & 
\\  & & & &   \\
$\theta =\mbox{}$&    &    & $  \frac{\theta_a}{C_v}(s -\mbox{\boldmath$\beta$}\cdot 
{\bf e}) $   &  $\frac{\theta_a}{C_p}(s - \bosy{\alpha}\cdot \bosy{\sigma})$
\\  & & & &   \\
\hline
\end{tabular}
}

\end{center}
\caption{Thermoelastic constitutive equations.  The left column indicates the 
 dependent  variable.  Each subsequent column corresponds to a set of constitutive equations 
 in terms of the independent variables in the first row.}\label{tableI}
\end{table}
%%%%%%%%%%%%%%%%%%%%%%%%%%%%%%%

The material constants are as follows:    
${\bf S}$ is  the fourth order tensor of isothermal compliances, with 
inverse ${\bf C}$, and corresponding adiabatic quantities are ${\bf S}_s$ and ${\bf C}_s$.  
 The symmetric second order tensor of thermal expansion coefficients is {\boldmath $\alpha$}, 
and the related tensor {\boldmath $\beta$} is defined as 
$\mbox{\boldmath$\beta$}  = {\bf C}      \mbox{\boldmath$\alpha$}$,
while the  quantities $C_p$ and $C_v$ are the heat capacities per unit volume 
at constant stress and strain respectively.  
The following relations can be verified from Table~\ref{tableI}: 
\beqa{rel}
 C_p/\theta_a = (C_v/\theta_a)+ \mbox{\boldmath$\alpha$}\cdot   {\bf C} 
\mbox{\boldmath$\alpha$},
\quad
 {\bf S} = {\bf S}_s + (\theta_a/C_p) \mbox{\boldmath$\alpha$}\otimes 
\mbox{\boldmath$\alpha$},
\quad
 {\bf C}  = {\bf C}_s - (\theta_a/C_v)\,  \mbox{\boldmath$\beta$} \otimes 
\mbox{\boldmath$\beta$}.
\eeqa 
A word about notation: 
 $ \bosy{\alpha}\cdot  \bosy{\sigma}= $tr$(\bosy{\alpha \sigma})$ is a scalar, while $  
\bosy{\alpha}\otimes \bosy{\alpha}$ is a fourth order tensor.

The constitutive relations in Table~\ref{tableI} follow from the standard thermodynamic 
relations  
\beqa{ther}
&{d}U = \bosy{\sigma}\cdot  {d}{\bf e} + \theta_a {d}s, \quad 
{d}F = \bosy{\sigma}\cdot  {d}{\bf e} - s_a {d}\theta , &
\\ && \nonumber \\
&{d}G = - {\bf e}\cdot  {d}\bosy{\sigma} - s_a {d}\theta, \quad 
{d}H = - {\bf e}\cdot  {d}\bosy{\sigma} + \theta_a {d}s, &
\eeqa
 where $U$, $F$, $G$ and $H$ are, respectively,  internal energy, Helmholtz free energy, 
 Gibbs free energy and enthalpy, all per unit volume.  These are related by the standard 
connections 
 $U= F+T S  = H + \bosy{\sigma}\cdot  {\bf e}= G +T S + \bosy{\sigma}\cdot  {\bf e}$, where 
 here, $T$ and $S$ are the absolute temperature and entropy,  $T = \theta_a +\theta$, $S= 
 s_a+s$.   The energy densities can be expressed, in the quadratic approximation that is used 
here, as
\beqa{ener}
&2U = {\bf e}\cdot  {\bf C}{\bf e}+ ( C_v /\theta_a)\,  \theta^2  , \quad 
2F = {\bf e}\cdot  {\bf C}_s{\bf e} - (\theta_a /C_v)\,  s^2 , &
\\ && \nonumber \\
&2G = -\bosy{\sigma}\cdot  {\bf S}_s\bosy{\sigma} - (\theta_a /C_p)\,  s^2  , \quad 
2H = - \bosy{\sigma}\cdot  {\bf S}\bosy{\sigma} + ( C_p /\theta_a)\,  \theta^2 , & \label{enerb}
\eeqa
 The constitutive relations  in Table~\ref{tableI} follow from  \rf{ther} and \rf{ener} 
combined 
with the basic definitions of the thermal expansion coefficients and the heat capacity,
\beq{defn}
\bosy{\alpha} = \left. {\partial {\bf e}\over \partial T }\right|_{\bosy{\sigma}},\quad  
\Delta Q_{p,v} = C_{p,v} \Delta T.
\eeq

\subsection{Thermoelastic relaxation governing equations}
 
  We first present the exact governing equations and then make appropriate asymptotic 
 approximations.  The motion is assumed to be caused by external forcing with no internal 
applied body forces or sources of heat. 
The heat flow in an elastic body is governed by the  energy balance
\beq{as4}
\theta_a \dot{s} + \mbox{div}\,{\bf q} = 0 , 
\eeq
where $\bf q$ is the heat flux.  
The equation for $s$ in terms of $\theta$ and $\bosy{\sigma}$ in Table~\ref{tableI} and 
\rf{as4} imply 
\beq{as5}
 C_p\dot{\theta} + \mbox{div}\,{\bf q}= -   \theta_a \, \mbox{\boldmath$\alpha$}\cdot  
\dot{\bosy{\sigma}}
\eeq
 Irreversibility is introduced by requiring the heat flux  to satisfy a generalized form of 
Fourier's relation \cite{Chand} 
\beq{as3}
{\bf q} + \tau_r \dot{\bf q}+ {\bf K} \nabla \theta = 0, 
\eeq
 where ${\bf K}$ is the positive definite thermal conductivity tensor, and $\tau_r$ is the 
 thermal relaxation time \cite{Chand}.  The Cattaneo--Vernotte heat flux equation \rf{as3} 
 includes the classical and more commonly used Kirchhoff law in the limit as $\tau_r 
 \rightarrow 0$.  The parameter $\tau_r$ is sometimes introduced to ensure finite speeds in 
 the theory \cite{lord67}.  
 Our objective is to solve the linear system of partial differential equations, 
 \rf{as5} and \rf{as3}, for $\theta$ as a function of the forcing in the right member of  
 \rf{as5}. The heat flux can be eliminated to give a single equation for~$\theta$, 
\beq{sin}
C_p\left( \dot{\theta} + \tau_r \ddot{\theta}\right ) - \mbox{div}\ {\bf K} \nabla \theta
 = -    \big( 1+ \tau_r\frac{\partial \, }{\partial t} \big) \theta_a\, 
\mbox{\boldmath$\alpha$}\cdot  \dot{\bosy{\sigma}}.
\eeq
A closed system of equations is obtained by applying the dynamic equilibrium condition 
 \beq{cs}
 \mbox{div} \bosy{\sigma} - \rho \ddot{\bf u} = 0,
\eeq
 where $\rho$ the mass per unit volume and  $\bf u$ is the elastic displacement vector, 
related to the strain via ${\bf e} = \big(
\nabla {\bf u}  + (\nabla {\bf u})^T\big)/2$.
 It is shown in Appendix A that closed-form solutions of the coupled system of equations 
 \rf{sin} and \rf{cs} are generally  feasible only under restricted conditions.  These 
 require, essentially, that the material must be elastically isotropic, which is too limiting 
for our purposes.

\section{Solution for arbitrary structures}\label{sec3}

\subsection{Asymptotic approximation}

  The key quantities are the positive definite compliance and stiffness differences $\Delta 
 {\bf S} \equiv   {\bf S} - {\bf S}_s$ and $\Delta {\bf C} \equiv   {\bf C}_s - {\bf C}$.  
 These determine the energy decrement between the final (isothermal) relaxed and initial 
 unrelaxed (adiabatic) states, and they follow from \rf{rel} as $\Delta {\bf S}  = 
 (\theta_a/C_p) \mbox{\boldmath$\alpha$}\mbox{\boldmath$\alpha$}^T$ and 
 $\Delta {\bf C} \equiv   {\bf C}_s - {\bf C} = (\theta_a/C_v) 
 \mbox{\boldmath$\beta$}\mbox{\boldmath$\beta$}^T$. 
 Zener's approach is based on a separation between the mechanics and the thermodynamics.    
 By assumption, the {\it total difference\/} between the relaxed and unrelaxed energies is 
 small. Specifically ${\Delta {\cal E}_0}/{{\cal E}_0} \ll 1 $, 
 where the mechanical energy ${\cal E}_0 = \frac12 \mbox{\boldmath$\sigma$}_0\cdot  {\bf 
 e}_0$ 
 is defined by ${\bf e}_0$ and $\mbox{\boldmath$\sigma$}_0$, which are related by the 
purely 
 mechanical equation $\mbox{\boldmath$\sigma$}_0  = {\bf C} {\bf e}_0$ (ignoring temperature 
 and entropy variations).  Thus,  
 ${\cal E}_0 = \frac12 \mbox{\boldmath$\sigma$}_0\cdot  {\bf S} \mbox{\boldmath$\sigma$}_0 $ 
 and the decrement may be defined  as  
\beq{as2}
 \Delta {\cal E}_0 = \mbox{\boldmath$\sigma$}_0\cdot  \Delta {\bf S} \, 
\mbox{\boldmath$\sigma$}_0
= (\theta_a/C_p) \big( 
\mbox{\boldmath$\alpha$}\cdot  \mbox{\boldmath$\sigma$}_0\big)^2 . 
\eeq
 Alternatively, $\Delta {\cal E}_0\approx {\bf e}_0\cdot  \Delta {\bf C} {\bf e}_0 = 
 (\theta_a/C_v) \big( \mbox{\boldmath$\beta$}\cdot  {\bf e}_0  \big)^2$, where the 
 approximation is due to the assumed purely mechanical relation between 
 $\mbox{\boldmath$\sigma$}_0 $ and $ {\bf e}_0$.
 The main point is that the relative change in energy between the unrelaxed and relaxed 
 states in either case is the same to leading order in  $\epsilon$, where the nondimensional 
 parameter governing TE damping is
\beq{eps}
\epsilon = {E\theta_a \alpha^2}/{C_p}. 
\eeq
 This definition of $\epsilon$ is chosen to equal the relative change in elastic moduli 
 $\Delta c/c_0$ for TE relaxation of a thin beam, equation \rf{q1a}.  It can 
also 
 be expressed as
\beq{epsa1}
\epsilon = \frac{1}{3}\,(1-2\nu )({C_p - C_v})/{C_p}, 
\eeq
 where $\nu$ is the isothermal Poisson's ratio.  Chadwick \cite{Chadwick62} employed a 
 slightly different nondimensional parameter (denoted here as  $\epsilon_c$ to distinguish it 
 from $\epsilon$) \beq{epsc}
\epsilon_c = \frac{1}{3}\left( \frac{1+\nu}{1-\nu} \right)\frac{C_p - C_v}{C_v}. 
\eeq
 It is clear that the nondimensional parameters are closely related, and in particular, of 
the same order of magnitude.

\subsection{Solution by projections for anisotropic systems}

 The coupled equations \rf{sin} and \rf{cs} are solved using a regular perturbation procedure 
in the 
 asymptotic parameter $\epsilon \ll 1$.    We will achieve the solution using a projection 
 method, similar to  Zener's approach.  A separation of variables reduces the problem to 
 coupled ordinary differential equations in time.  Anisotropy in the elastic material does 
 not permit a modal expansion with a common set of scalar eigenfunctions, the basis for 
 Zener's method, and the key to a  generalization of his method to the limited but important 
 case of isotropic solids \cite{Chadwick62}, see Appendix A.  However,  even in the case of 
 the exact solution obtained by Chadwick \cite{Chadwick62}, the interesting phenomena are 
 obtained by the leading order approximation to the complex-valued frequencies.  It therefore 
 makes more sense ultimately to proceed by a regular asymptotic approximation at the stage of 
 the coupled equations \rf{sin} and \rf{cs}.   In this approach we view them as decoupled to 
 leading order, whereby the elasticity problem is solved with no thermal effects.  That is, 
 we consider the elasticity as an  uncoupled but vital forcing term in the `thermal 
equation'~\rf{sin}.

 Thus, to leading order equation \rf{sin} gives an uncoupled equation for temperature 
with the stress entering on the right hand side as a forcing term. 
  It is important to note that the forcing in \rf{sin} is proportional to $\bosy{\alpha}\cdot  
 \dot{\bosy{\sigma}}$. When the thermal expansion tensor is isotropic, the forcing depends 
 upon the rate of hydrostatic stress, even when the material is elastically anisotropic.  
 Thus, it is the hydrostatic stress,  {\it not\/} strain, that governs the TE~loss.

 Further progress  is made using projections onto a set of eigenfunctions.   In fact, this is 
 similar to the method first proposed by Zener \cite{Zener}, which treated a simpler 
 decoupled thermoelasticity problem.   We  first discuss the generalization of Zener's method 
 to  \rf{sin}  as it allows us to determine the final answer in a form similar to the 
 familiar and classical result of Zener for an elastic beam in flexure.  We will later 
 compare the general solution with  direct solutions for particular configurations. 
 Assume the temperature can be represented~as 
\beq{as6}
\theta ({\bf x}, t) = \sum\limits_{n=0}^\infty \theta_n (t) \phi_n({\bf x}), 
\eeq
where ${\bf x} = (x,y,z) = (x_1,x_2,x_3)$ and the eigenvalues $\tau_n$ and eigenfunctions $\phi_n$ satisfy 
\beq{as7}
 C_p^{-1} \mbox{div}\, {\bf K} \nabla \phi_n + \tau_n^{-1} \phi_n = 0, 
 \quad n=0,\, 1, \, 2, \ldots , 
 \eeq
 plus appropriate boundary conditions (for example,  no flux).   The amplitudes solve
  \beq{as8}
\dot{\theta}_n +  \tau_r \ddot{\theta}_n + \tau_n^{-1} \theta_n = 
 -   \left( 1+ \tau_r\frac{\partial \, }{\partial t} \right)  \frac{\theta_a}{C_p} \langle 
\phi_n , \,   \bosy{\alpha}\cdot  \dot{\bosy{\sigma}} \rangle 
\eeq
 where the brackets indicate the inner product 
$\langle f , g \rangle = \int dV f({\bf x})\,g({\bf x})$.
 
 \subsection{A general result for energy loss}
 
  Before considering applications of  \rf{as6}--\rf{as8} to particular structures we 
first derive a general result for TE dissipation. 
  A measure of local structural damping may be defined in terms of the local relative loss in 
energy per cycle.  
 The rate of change of local mechanical energy per unit volume  is 
$\dot{{\cal E}} = \mbox{\boldmath$\sigma$}\cdot  \dot{\bf e}$. 
 We assume periodic oscillation for $\bosy{\sigma}$ and {\bf e}  and determine the  loss in 
 the mechanical energy through the coupling to irreversible thermal process,  $\theta$, also 
 periodic.  Using the relation for {\bf e} in terms of $\bosy{\sigma}$  and $\theta$ in 
Table~\ref{tableI}, gives 
\beq{a10a}
{\dot{\cal E}}   = \mbox{\boldmath$\sigma$}\cdot  {\bf S} \dot{\bosy{\sigma}}
+ { \bosy{\sigma}\cdot \bosy{\alpha} \, \dot{\theta} }, 
\eeq
Taking the average over a cycle,  and  
 using $\overline{f\dot{f}}=0$ where the overbar indicates a time average,  implies that the 
local irreversible energy loss rate per unit volume is 
\beq{a10}
 \overline{\dot{\cal E}({\bf x})}  =    -\, \overline{\theta ({\bf x}, t)\, \bosy{\alpha}\cdot 
 \dot{\bosy{\sigma}}({\bf x}, t) } = \sum_{n = 0}^{\infty}\phi_n ({\bf x}) \, \overline{\theta_n (t) 
\bosy{\alpha}\cdot \dot{\bosy{\sigma}}({\bf x},t) }. 
\eeq
 The total power dissipated is thus
\begin{eqnarray}
 {\Pi_d} & = & - \int dV \overline{\dot{\cal E}({\bf x})} = - \sum_{n = 0}^{\infty} 
 \overline{\theta_n (t) \langle \phi_n , \,  \bosy{\alpha}\cdot \dot{\bosy{\sigma}} \rangle } 
 \nonumber \\
\label{a11}
 & = &  \sum_{n = 0}^{\infty} \frac{\omega^2 \tau_n}{(1- \omega^2 \tau_r\tau_n)^2 + \omega^2 
\tau_n^2}\, \frac{ \theta_a}{C_p}\, 
 \overline{\langle \phi_n , \,  \bosy{\alpha}\cdot  \bosy{\sigma} \rangle^2 }
\end{eqnarray}
 where we have assumed periodic motion of period $2\pi/\omega$ and used \rf{as8} to 
express the time harmonic temperature field in terms of the stress.

 The loss factor is then given by $ Q^{-1} \approx \tan \delta = \Delta {\cal E}/(2\pi {\cal 
 E}_0) $, where $\Delta {\cal E} = \overline{\Pi_d} \, 2\pi/\omega$ is the loss per cycle and 
 ${\cal E}_0$ is the total energy of oscillation. 
We find
\beq{a13}
Q^{-1} ={\theta_a \over C_p {\cal E}_0}\sum\limits_{n=0}^\infty 
 \frac{\omega \tau_n}{(1- \omega^2 \tau_r\tau_n)^2 + \omega^2 \tau_n^2}\, 
\overline{\langle \phi_n , \, \mbox{\boldmath$\alpha$}\cdot  
\mbox{\boldmath$\sigma$}\rangle^2 }. 
\eeq
 This provides a general formula for the TE loss in terms of the inhomogeneous 
 stress.  The $Q$ of a particular mode may be straightforwardly obtained by integrating  
 (\ref{a13}) over the volume of the oscillator and dividing by the total energy.  This 
 equation may alternatively be expressed in terms of the inhomogeneous strain, however, we  
find the stress formulation more convenient.

 Equation \rf{a13} is one of the main results of the paper, as it provides a means to compute 
TE dissipation given a solution in terms of the inhomogeneous stress.

 We remark on  the summation in \rf{a13}.  If the first term in the infinite sum is dominant, 
 as is often the case \cite{Zener}, the sum can be truncated after only one term ($n=0$).  
 This gives a result very similar to  \rf{q1} except that the simple Lorentzian in the latter 
is replaced by the generalized Lorentzian amplitude 
\beq{a}
A (\omega\tau) = \frac{\omega \tau}{(1- \omega^2 \tau_r\tau)^2 + \omega^2 \tau^2}. 
\eeq
 Of course, this reduces to the classical Lorentzian when $\tau_r=0$, which has a maximum as 
 a function of $\omega$ when $\omega \tau = 1$.   It is worth describing  the properties of 
 this generalized  Lorentzian, in particular  how $\tau_r$ influences the maximum.   For 
every $\tau_r \ge 0$, $A$  
 has a single maximum at a unique value of $\omega = \omega^*$ defined by 
\beq{aaa}
\omega^* \tau = \big[ (1-4r + 16r^2)^{1/2} -1 + 2r\big]^{1/2}/(r\sqrt{6})
\quad 
\mbox{where} \quad r = {\tau_r}/{\tau }.   
\eeq
 Furthermore,  $\omega^* \tau \ge 1$ for a restricted range of $\tau_r$.  Specifically, 
  $1\le \omega^* \tau \le \sqrt{4/3}$  for  $0\le \tau_r \le 2\tau/3$, with $\omega^* \tau $ 
 equal to unity at the two extremes ($\tau_r = 0$ and $\tau_r = 2\tau/3$) and $\omega^* \tau 
 = \sqrt{4/3}$ for  $\tau_r =\tau/4$.   Conversely, $\omega^* \tau <  1$ for $\tau_r > 
 2\tau/3$.  
  In particular, for relatively large $\tau_r \gg \tau$, the maximum is at $\omega^* (\tau 
 \tau_r)^{1/2} = 1$.  The value of $A$ at the maximum, $A_{max}$, increases monotonically 
 from $A_{max}=1/2$ when there is zero thermal relaxation,  $\tau_r = 0$,  to $A_{max} \approx 
 (\tau_r/ \tau)^{1/2}$ for $\tau_r \gg \tau$.

\section{Thermoelastically damped orthotropic thin plates}\label{sec4}

\subsection{Thin plate dynamics}

 We consider plates that are orthotropic with  axes of symmetry coincident with the 
 coordinate axes.  Assume, with no loss in generality, that  the thermal expansion tensor is 
 diagonal $\bosy{\alpha} = $diag$(\alpha_1 ,\, \alpha_2 , \, \alpha_3)$.  By virtue of the 
 thin plate configuration we can ignore the normal stress $\sigma_{zz}$, and employing the 
 Kirchhoff approximation for the deformation, we have 
\beq{k1}
\bosy{\alpha}\cdot \bosy{\sigma} = \alpha_1\sigma_{xx}+\alpha_2\sigma_{yy}
= \frac{ z}{1-\nu_{12}\nu_{21}} \big[ (1+\nu_{21})E_1\, \alpha_1 \kappa_{xx}   +  
(1+\nu_{12})E_2\, \alpha_2 \kappa_{yy}\big], 
\eeq
 where $\bosy{\kappa}$ is the curvature tensor with components $\kappa_{xx}$, $\kappa_{yy}$ 
 and $\kappa_{xy}$.  The curvature  is related to the transverse deflection of the 
centre-plane, $w(x,y)$, by 
\beq{cu}
{\kappa_{ij}} = - {\partial^2 w(x,y) \over \partial x_i \partial x_j} \, . 
\eeq
 The quantity $\nu_{ij}$   is the Poisson ratio for strain in the $j$-direction caused by 
 stress in the $i$-direction, and the two Poisson's ratios satisfy $\nu_{21}E_1 
=\nu_{12}E_2$.  
 The instantaneous potential energy density per unit area of a plate in flexure is
\beq{ed}
{\cal E}_{PE}=\frac{I}{2}\left\{
\frac{1}{1-\nu_{12}\nu_{21}}\big( 
  E_1\kappa_{xx}^2 +E_2\kappa_{yy}^2 + 2\nu_{21}E_1\kappa_{xx}\kappa_{yy}\big) + 4\mu 
\kappa_{xy}^2\right\},
 \quad
\eeq
where  $\mu $ is the in-plane shear modulus and $I\equiv \langle z, \,  z \rangle =h^3/12$.

\subsection{Asymptotic solution by projections for  thin plates}

 Our purpose here is to obtain a general solution for the TE loss based on the assumption 
 that the transverse diffusion of heat can be ignored.  This assumption is examined in 
 Section \ref{sec5} where it is shown that provided the assumptions of thin plate theory and 
 classical heat transport are satisfied, the transverse heat flow gives rise only to small 
 corrections to the TE loss.

 The governing equation for the non-equilibrium temperature field is obtained by ignoring the 
transverse heat flow terms in  (\ref{sin}),
\beq{doug1}
 C_p\left( \dot{\theta} + \tau_r \ddot{\theta}\right ) - \ K_3 {\partial^2 \theta \over 
\partial z^2}
 = - \left( 1+ \tau_r\frac{\partial \, }{\partial t} \right) \theta_a\, 
\mbox{\boldmath$\alpha$}\cdot  \dot{\bosy{\sigma}}
\eeq
 where the dependence of the local temperature on position is governed solely by the  
${\bf x}$-dependence of the prescribed stress field $\sigma$ where now ${\bf x}=(x,y)=(x_1,x_2)$.  Also, $K_3$ is the through-thickness 
 element of the thermal conductivity tensor, which in keeping with the general orthotropic 
formulation, is ${\bf K} = {\rm diag}\,(K_1,\, K_2,\, K_3)$.

 The analysis leading to   (\ref{a11}) for the power dissipated by TE effects can be repeated 
 with the proviso that the eigenfunctions $\phi_n$ of the heat equation are now functions 
only 
 of $z$, the thickness direction, and inner products should in this case be interpreted 
 accordingly as $\langle f, g \rangle = \int dz f(z)\, g(z)$.  Hence the analogue to 
(\ref{a11}) 
 refers not to the total power lost, but instead to the rate of energy loss per unit area at 
 position ${\bosy x}$;
\begin{equation}
\label{doug2}
   \dot {\cal E}({\bosy x}) = - \, \sum_{n = 0}^{\infty} \frac{\omega^2 \tau_n}{(1- \omega^2 
\tau_r\tau_n)^2 + \omega^2 \tau_n^2}\, \frac{ \theta_a}{C_p}\, 
 \overline{\langle \phi_n(z) , \,  \bosy{\alpha}\cdot  \bosy{\sigma}({\bosy x}, z) \rangle^2 }
\end{equation}

The temperature modes of interest are antisymmetric about the mid-plane \cite{Zener} with 
\beq{mod}
\phi_n = ({2}/{h})^{1/2}  \sin (2n+1)\frac{\pi z}{h},\quad 
\tau_n = (2n+1)^{-2}\, \frac{h^2C_p}{\pi^2 K_3},\quad n=0,1,2,\ldots . 
\eeq
  Temperature modes that are symmetric in $z$ have zero coupling to   flexural stress 
 components, since they are antisymmetric, and also to any membrane stresses, which are 
constant 
 across the thickness.  It is evident  from  \rf{doug2} and \rf{k1} that the thermal loss in 
flexure  depends upon the quantities \cite{Zener}
\beq{inn}
f_n  \equiv  \langle \phi_n , \,  z \rangle^2 /   \langle z, \,  z \rangle
 =96/[(2n+1)\pi]^4.
% \nonumber \\
%  &=&  \left( \big( \frac{2}{h}\big)^{1/2} \int\limits_{-h/2}^{h/2} z \sin (2n+1)\frac{\pi 
% z}{h}\, \mbox{d}z \right)^2 / \int\limits_{-h/2}^{h/2} z^2 \, \mbox{d} z
% \nonumber \\
% &=& \frac{96}{\pi^4 (2n+1)^4}\, . 
\eeq 
 Combining  \rf{k1}, \rf{doug2}, and \rf{inn}, we obtain
\beq{f1}
 { \dot {\cal E}({\bosy x}) }  =  - \, {\cal E}_{Diss}( \bosy{\kappa} (\bosy{x}))\, 
 \frac{\bar{E} \theta_a \bar{\alpha}^2}{C_p}
 \sum\limits_{n=0}^\infty f_n
 \frac{\omega^2 \tau_n}{(1- \omega^2 \tau_r\tau_n)^2 + \omega^2 \tau_n^2}
\eeq
 where $\bar E$ $=(E_1 + E_2)/2$ is the average Young's modulus, $\bar{\alpha} = (\alpha_1 + 
 \alpha_2)/2$, $\bosy \kappa$ is the curvature tensor, and the `dissipation energy density'  
(per unit area) ${\cal E}_{Diss}$ is given by
\beq{doug3}
 {{\cal E}_{Diss}}( \bosy{\kappa} (\bosy{x}))= \frac{I}{\bar{E}\bar{\alpha}^2 }
 \left[  \frac{ (1+\nu_{21})E_1\alpha_2\,\kappa_{xx}   + (1+\nu_{12})E_2\alpha_2\, 
\kappa_{yy}} {1-\nu_{12}\nu_{21}}\right]^2.  
 \eeq
 Equation \rf{f1} is a key result pertaining to TE dissipation in structures which can be 
 modelled as thin plates.  Unlike most previous results, the predicted dissipation is 
inhomogeneous.

 The local TE dissipation depends on position via the dependence on the local curvature 
 tensor $\bosy \kappa (x)$.  This aspect may be explored by defining the quantity 
 $p(\bosy{\kappa} ) = {{\cal E}_{Diss}}( \bosy{\kappa}) / {\cal E}_0(\bosy{\kappa} )$, which 
 gives a measure of the local TE energy dissipation relative to the local deformation energy.  
 Expressing the total energy as twice the average potential energy by virtue of the virial 
 theorem we find, 
\beq{f3}
p(\bosy{\kappa} ) = 
 \frac{ \left[{\bar E}\bar{\alpha}^2(1-\nu_{12}\nu_{21})\right]^{-1}\, \big[  
(1+\nu_{21})E_1\alpha_1\,\kappa_{xx}   + (1+\nu_{12})E_2\alpha_2\, \kappa_{yy}\big]^2}
  {E_1\kappa_{xx}^2 +E_2\kappa_{yy}^2 + 2 \nu_{21}E_1\kappa_{xx}\kappa_{yy} + 4\mu 
(1-\nu_{12}\nu_{21})
 \kappa_{xy}^2 }.
 \quad
\eeq

The parameter $p$ simplifies for  materials of cubic symmetry, such as silicon, with $E_1 = 
E_2 = \bar E  \equiv E$, $\alpha_1=\alpha_2$ and $\nu_{12}= \nu_{21}\equiv \nu$, and hence
\beq{f4}
p(\bosy{\kappa} ) = \left( \frac{1+\nu}{1- \nu}\right)
\frac{ \big(  \kappa_{xx}   +  \kappa_{yy}\big)^2}
  {\kappa_{xx}^2 + \kappa_{yy}^2 + 2\nu \kappa_{xx}\kappa_{yy} + 4E^{-1}\mu (1-\nu^2) 
\kappa_{xy}^2 }.
 \eeq
For isotropic materials, $E=2\mu (1+\nu)$, and $p(\bosy{\kappa} )$ becomes
\beq{f5}
p(\bosy{\kappa} ) =  \left ( \frac{1+\nu}{1- \nu}\right )
{{ (\mbox{tr}\bosy{\kappa})^2 } \over
   { (\mbox{tr} \bosy{\kappa})^2 - 2(1-\nu ) \det\bosy{\kappa} } }.
 \eeq
 In this case, $p(\bosy{\kappa} )$ depends upon the two principal invariants of the 
curvature: 
 $\mbox{tr} \bosy{\kappa} =  \kappa_{xx}   +  \kappa_{yy}$ and $\det\bosy{\kappa} =  
\kappa_{xx}\kappa_{yy}- \kappa_{xy}^2$. 
Let $\kappa_1$ and $\kappa_2$ be the two principal curvatures, satisfying 
$\kappa_1+\kappa_2=\mbox{tr}\bosy{\kappa}$ and $\kappa_1\kappa_2=\det\bosy{\kappa}$, then 
\beq{f6}
p(\bosy{\kappa} ) = \left( \frac{1+\nu}{1- \nu}\right)
\frac{ \big(  \kappa_1+\kappa_2\big)^2}
 {\big(  \kappa_1+\kappa_2\big)^2 - 2(1-\nu) \kappa_1\kappa_2 }.
\eeq
Thus, 
\beq{f7}
0\le p(\bosy{\kappa} ) \le {2}/(1- \nu),
\eeq
 with $p(\bosy{\kappa} )=0$ at locations where the plate is locally saddle-shaped, 
 $\kappa_1=-\kappa_2$, and $p$ achieves its maximum value at points where  it is locally 
 spherical, $\kappa_1=\kappa_2$.

The   bounds \rf{f7} also apply  to materials of cubic anisotropy, and occur in the same 
circumstances as for the isotropic material, as can be verified from  \rf{f4}.

\subsection{TE loss factors for flexural modes and waves}

 The loss factor of a particular flexural mode may be predicted via (\ref{f1}) once the 
 displacement field $w(x)$ of the mode is known.  The curvature tensor is first evaluated via 
 the standard relation  (\ref{cu}).  The total energy lost from the mode per cycle is then 
 calculated by integrating  (\ref{f1}) over the volume of the plate and time averaging.  In 
 the most common situation, the displacement field will be evaluated in frequency space as a 
 complex quantity, and the time average is obtained in the usual way as $\overline {f(t) 
 g(t)}=\frac{1}{2}{\rm Re}\,[\tilde{f}(\omega ) \tilde{g}^*(\omega )]$, where $\tilde{f}$ is the Fourier transform.  Thus the TE dissipation for mode 
$\sigma$ is  given by
\begin{equation}
\label{dp2}
Q_\sigma^{-1} =
 \frac{\bar E \theta_a\bar{\alpha}^2}{C_p}
 \sum\limits_{n=0}^\infty f_n
  \frac{\omega \tau_n}{(1- \omega^2 \tau_r\tau_n)^2 + \omega^2 \tau_n^2}\, \left[ {\int dA  
~{ {{\cal E}_{Diss}}( \bosy{\kappa}) } \over {\cal E}_{0\sigma} }\right]
\end{equation}
 where ${\cal E}_{0\sigma}
= \int dA \, ( { {\cal E}_{KE}} + { {\cal E}_{PE}})$  is the total energy of mode $\sigma$.

 The modal energy ${\cal E}_{0\sigma}$ may be computed as either twice the average kinetic 
 energy of the system or twice the average potential energy of the system, whichever is more 
 convenient.  The kinetic energy is often preferable when using experimental data because the 
 second derivatives appearing in the curvature tensor can be ill behaved in the presence of 
noise.

 The expression for the TE loss factor given above is closely related to the loss factor 
 given by Zener for a simple beam in flexure.  Zener gave the result (see  \rf{q1a}),
\begin{equation}
\label{Zener}
Q_{{\rm Zener}}^{-1} =
 \frac{ E \theta_a\alpha^2}{C_p}
 \frac{\omega \tau_0}{1 + \omega^2 \tau_0^2}\, 
\end{equation}
 where we have included only the first term in the infinite sum.  Our results may thus be 
interpreted as
\beq{mpf}
 Q_\sigma^{-1} =  {\int dA\,  {{{\cal E}_{Diss}}( \bosy{\kappa}) } \over {\cal E}_{0\sigma} 
}\, Q_{{\rm Zener}}^{-1} = p_\sigma \, Q_{{\rm Zener}}^{-1} 
\eeq
 where the quantity $p_\sigma$ was called the modal participation factor (MPF) in the 
isotropic 
 case studied in~\cite{Photiadis02}.  The MPF is evidently closely related to the 
 quantity $p(\bosy{\kappa} )$ analyzed above.

 The MPF simplifies considerably for an isotropic medium.  Using similar manipulations as 
employed above, we find
\begin{equation}
\label{dp3}
p_\sigma = {I E \over {\cal E}_0 (1-\nu )^2} \int dA ~\overline{ (\mbox{tr} \bosy{\kappa} )^2 }
\end{equation}
 a result which differs from that given in  \cite{Photiadis02} by a factor of $(1-\nu 
 )^{-2}$.  The reason for the difference is that it was assumed in  \cite{Photiadis02} 
that  flexural energy would be dissipated at the rate predicted by Zener for a beam.
However, a plate undergoes more compressive stress than a beam  
and thus dissipates more energy accordingly.

 It is not difficult to show, based on \rf{dp3}, that the MPF for travelling flexural waves 
in 
an isotropic plate is 
\begin{equation}
\label{dp5}
p_\sigma = {1 +\nu \over 1-\nu }\, \quad \mbox{for a flexural wave}. 
\end{equation}
 This shows the difference between the TE dissipation of a cantilever beam ($p_\sigma = 1$) 
 versus the corresponding vibration of a plate.
 The identity \rf{dp5}  may be obtained by explicit substitution of a flexural waves 
 solution, or more simply, as follows.  Integrating by parts and using the governing plate 
 equation $EI(1-\nu^2)^{-1}\nabla^4 w -  \rho h \omega^2 w =0$, gives 
\beq{gives}
\int dA ~\overline{ (\mbox{tr} \bosy{\kappa} )^2 } =
\int dA ~\overline{  w \nabla^4 w} = \frac{2(1-\nu^2)}{IE}
\int dA ~\overline{ {\cal E}_{KE} }= \frac{(1-\nu^2)}{IE}
\, {\cal E}_{0}.
\eeq
 The last equality employs the expression for the potential energy,  (\ref{ed}).  Also, it is 
 assumed that the plate boundary conditions may be ignored in the above integration, which is 
 true for a travelling wave in a plate of `infinite' extent.

\section{Effective thin plate equations and flexural waves}\label{sec5}

 The general theory is now applied to thin plates in flexure with the goal of deriving 
 general governing equations similar to the classical Kirchhoff thin plate equations.  Our 
 objective is to provide an alternative means of calculating $p_\sigma$ for flexural waves, 
 and also to examine the range of validity of our approximation in ignoring lateral thermal 
 diffusion.

 The equation for $\theta$, \rf{sin}, becomes for time harmonic motion ($e^{-i\omega t}$ 
 assumed)
\beq{sin2b}
 \frac{\partial^2 \theta}{\partial z^2} + k^2 \theta
= -   k^2 (\theta_a/C_p) \bosy{\alpha}\cdot  \bosy{\sigma} - \left( \frac{K_1}{K_3}
 \frac{\partial^2 \theta}{\partial x^2} + \frac{K_2}{K_3}\frac{\partial^2 \theta}{\partial 
y^2}\right). 
\eeq
where  ${\bf K} = {\rm diag}\,(K_1,\, K_2,\, K_3)$ and
\beq{sin3}
 k^2 = i\omega (1 - i\omega \tau_r) {C_p}/{K_3}.
\eeq
 If the stress term on the right hand side of \rf{sin2b}  is weakly dependent on $x$ and $y$, 
 then we may argue that $\theta$ inherits the same weak dependence. Provided this dependence 
 on transverse position is uniform, the case of simple plane wave, the final term may be 
 combined with the $k^2 \theta$ term,  and hence interpreted as modifying the thermal 
 diffusion rate.  Here, we ignore this  and simplify  \rf{sin2b} to 
\beq{sin2}
 \frac{\partial^2 \theta}{\partial z^2} + k^2 \theta
= -   k^2 (\theta_a/C_p) \bosy{\alpha}\cdot  \bosy{\sigma}.
\eeq
 The importance of the simplification \rf{sin2} at this stage is that it allows us to derive 
 a set of effective plate equations in which the TE damping appears directly.  
 This approach follows on that of Alblas \cite{Alblas81} and of Lifshitz and Roukes 
 \cite{Lifshitz00} who derived the effective equation governing the motion of a 
 thermoelastically damped beam.

 Assuming zero flux conditions at $z=\pm h/2$, and noting that the right member of  \rf{sin2} 
 is proportional to $z$ in flexure,  $ \bosy{\sigma} =  \langle z,\,
 \bosy{\sigma}\rangle I^{-1} z$,  we find that the solution~is
% for the temperature is 
\beq{sin4}
\theta = -   {\theta_a\over I C_p} \langle z,\, \bosy{\alpha}\cdot  \bosy{\sigma}\rangle\, 
\left( z - \frac{\sin kz}{k \cos({kh}/{2})} \right).
\eeq
 The thermally perturbed stress is therefore, using the expression for entropy in terms of 
temperature and strain from Table~\ref{tableI}, 
\beq{th1}
\bosy{\sigma} = {\bf C}{\bf e} + {\theta_a\over I C_p} \langle z,\, \bosy{\alpha}\cdot  \bosy{\sigma}_0\rangle\, 
\left( z - \frac{\sin kz}{k \cos({kh}/{2})} \right)\, \bosy{\beta}\, .
\eeq
 The stress in the absence of  thermal effects,  $\bosy{\sigma}_0$, is proportional to $z$.   
 In order to apply  \rf{th1} to the thin plate the standard plane stress conditions must be 
 enforced.  We consider an orthotropic plate with a symmetry plane coincident with the 
 neutral plane $(z=0)$, for which the standard stress/strain relations for plane-stress 
are 
\beqa{ps1}
\left[\ba{c} 
\sigma_{xx}^{(0)} \\ \\ 
\sigma_{yy}^{(0)} \\ \\ 
\sigma_{xy}^{(0)}
\ea \right] = 
\left[\ba{ccc} 
\frac{E_1}{1-\nu_{12}\nu_{21}} & \frac{\nu_{21}E_1}{1-\nu_{12}\nu_{21}} & 0 \\
& & \\
\frac{\nu_{12}E_2}{1-\nu_{12}\nu_{21}} & \frac{E_2}{1-\nu_{12}\nu_{21}} & 0 \\
& & \\
0 & 0 & 2\mu 
\ea \right]
\left[\ba{c} 
e_{xx}^{(0)} \\ \\ 
e_{yy}^{(0)} \\ \\ 
e_{xy}^{(0)}
\ea \right]
\eeqa
 where $\sigma_{xx}^{(0)}$,  $e_{xx}^{(0)}$, etc. are the stresses and strains in the absence 
of the thermoelastic damping. 
Based on  \rf{th1} this implies that the in-plane TE  stresses are  
\beqa{plst1}
 \sigma_{xx} &=& \sigma_{xx}^{(0)}+ \frac{(\alpha_1+\alpha_2\nu_{21})\theta_a 
 E_1}{(1-\nu_{12}\nu_{21}) C_p I} \langle z,\, \alpha_1\sigma_{xx}^{(0)}+\alpha_2 
 \sigma_{yy}^{(0)}\rangle\, 
\bigg( z - \frac{\sin kz}{k \cos({kh}/{2})} \bigg),
\nonumber \\ 
 \sigma_{yy} &=& \sigma_{yy}^{(0)}+ \frac{(\alpha_2+\alpha_1\nu_{12})\theta_a 
 E_2}{(1-\nu_{12}\nu_{21}) C_p I} \langle z,\, 
 \alpha_1\sigma_{xx}^{(0)}+\alpha_2\sigma_{yy}^{(0)}\rangle\, 
 \bigg( z - \frac{\sin kz}{k \cos({kh}/{2})} \bigg), 
\\
\sigma_{xy} &=& \sigma_{xy}^{(0)}. \nonumber 
\eeqa
  This provides the variation of  stress through the thickness due to the temperature 
 variation.  The moments are found by taking the first moment of the stresses through the 
 plate thickness, leading to 
\beqa{plst2}
 M_{xx} &=& M_{xx}^{(0)}+ \frac{(\alpha_1+\alpha_2\nu_{21}) \theta_a 
E_1}{(1-\nu_{12}\nu_{21}) C_p }  f(kh)\, \big( \alpha_1M_{xx}^{(0)}+\alpha_2M_{yy}^{(0)}\big),
\nonumber \\ 
 M_{yy} &=& M_{yy}^{(0)}+ \frac{(\alpha_2+\alpha_1\nu_{12}) \theta_a 
E_2}{(1-\nu_{12}\nu_{21}) C_p }  f(kh)\, \big(  \alpha_1M_{xx}^{(0)}+\alpha_2M_{yy}^{(0)}\big), 
\\
M_{xy} &=& M_{xy}^{(0)}, \nonumber 
\eeqa
where $M_{xx}^{(0)} = \langle z,\, \sigma_{xx}^{(0)}\rangle$, etc., and the function $f$ is 
\beq{aa3}
f(\zeta) = 1+ \frac{24}{\zeta^3}\bigg[ \frac{\zeta}{2} - \tan \frac{\zeta}{2} \bigg]. 
\eeq
 The  standard moment-curvature relations follow from \rf{ps1} and the Kirchhoff assumption 
as
\beqa{ps2}
\left[\ba{c} 
M_{xx}^{(0)} \\ \\ 
M_{yy}^{(0)} \\ \\ 
M_{xy}^{(0)}
\ea \right] = I 
\left[\ba{ccc} 
\frac{E_1}{1-\nu_{12}\nu_{21}} & \frac{\nu_{21}E_1}{1-\nu_{12}\nu_{21}} & 0 \\
& & \\
\frac{\nu_{12}E_2}{1-\nu_{12}\nu_{21}} & \frac{E_2}{1-\nu_{12}\nu_{21}} & 0 \\
& & \\
0 & 0 & 2\mu 
\ea \right]
\left[\ba{c} 
\kappa_{x}^{(0)} \\ \\ 
\kappa_{y}^{(0)} \\ \\ 
\kappa_{xy}^{(0)} 
\ea \right].  
\eeqa

The governing equation for the thin plate is  
\beq{plst3}
 \frac{\partial^2 M_{xx}}{\partial x^2} + 2 \frac{\partial^2 M_{xy}}{\partial x\partial y}
+ \frac{\partial^2 M_{yy}}{\partial y^2} + \rho h \omega^2 w = 0. 
\eeq
Thus, we obtain the effective thin plate equation for $w$, 
\beqa{ps3}
&&  - \frac{I}{1-\nu_{12}\nu_{21}}\bigg\{
E_1w_{xxxx}+E_2w_{yyyy} + 2 \big[ 2\mu (1-\nu_{12}\nu_{21})+ \nu_{21}E_1\big] w_{xxyy} \bigg\}
\nonumber \\ 
&& \mbox{}- \frac{  f(kh)\theta_a I }{(1-\nu_{12}\nu_{21})^2C_p }
 \bigg\{ (\alpha_1+\alpha_2\nu_{21})^2E_1^2w_{xxxx} + 
2(\alpha_1+\alpha_2\nu_{21})(\alpha_2+\alpha_1\nu_{12})E_1E_2w_{xxyy} 
\nonumber \\ 
&&      
\quad \mbox{}
+ (\alpha_2 +\alpha_1 \nu_{12})^2E_2^2w_{yyyy}\bigg\}\, + \rho h \omega^2 w = 0. 
\eeqa
For a plate made of material with cubic symmetry, this reduces to 
\beq{rt}
 \left[1 + \left(\frac{1+\nu}{1-\nu}\right)\frac{E\theta_a \alpha^2}{C_p }  f(kh) 
\right] \nabla^4 w
+ 4(1-\nu)\left[ (1+\nu)\frac{\mu}{E} - \frac{1}{2}\right] w_{xxyy} 
  - \frac{1-\nu^2}{EI}\rho h \omega^2  w = 0. \,\,
\eeq
 The second term vanishes for isotropic materials, in which case we may combine the damping 
 with the plate stiffness to get an equation in the standard form, 
\beq{plst5}
  {D}\nabla^4 w
- {\rho h} \omega^2 \,  w = 0, 
\eeq
where the TE loss is now contained in the complex-valued flexural stiffness  
\beq{plst6}
 D =  \frac{EI}{1-\nu^2} \left\{
1 +  \left(\frac{1+\nu}{1-\nu}\right)\frac{E\theta_a \alpha^2}{C_p } f(kh) 
\right\}.
\eeq
 In the more general orthotropic situation, the TE damping is inhomogeneous, as 
in 
 \rf{ps3}, and cannot be interpreted in terms of a single frequency-dependent complex modulus 
 of elasticity.

 Our   equation \rf{plst5}, a generalization of Zener's results for a beam 
 \cite{Zener37}, gives a homogeneous damping as a result of assuming a uniform plane wave 
 vibration.  This provides some guidance as to the loss factors of vibrating thin plate 
 structures but finite structures will support resonant modes that consist of a variety of 
 wavenumbers which can interfere with one another, and thus the actual loss factor for a 
 particular mode may differ significantly from this value.  A particularly salient example is 
 the case of a twisting mode for which the source field for temperature fluctuations, 
$\mbox{tr} \bosy \sigma$ (for isotropic $\bosy{\sigma}$), is very small, and the resulting values of 
 $Q^{-1}$ may be orders of magnitude smaller than the result given in~\rf{plst5}.

 Corresponding results for flexural waves in circular rods are in Appendix C.  In particular 
 we note that the function $f$ takes on a different form from that for a plate.

\subsection{Dispersion relation including in-plane variation}

 It is now relatively straightforward to revise the  analysis of Section \ref{sec4} to 
 include in-plane variation in both the stress and the temperature.   We begin by assuming 
 that  all field variables possess in-plane dependence $e^{i\zeta x}$ with wavenumber 
 $\zeta$.  Then  \rf{sin2b} becomes 
\beq{sin21}
 \frac{\partial^2 \theta}{\partial z^2} + \gamma^2 \theta
= -   k^2 (\theta_a/ C_p) \bosy{\alpha}\cdot  \bosy{\sigma}, 
\eeq
where $\gamma =( k^2 -\zeta^2K_1/K_3)^{1/2}$ 
and $k$ is defined in~\rf{sin3}.   Solving as before, we find that the temperature is 
\beq{sin22}
\theta = -    \langle z,\, \bosy{\alpha}\cdot  \bosy{\sigma}\rangle\, 
 \frac{\theta_a}{I C_p} \, \frac{k^2}{\gamma^2}\, 
\bigg( z - \frac{\sin  \gamma z}{\gamma \cos({\gamma h}/{2})} \bigg), 
\eeq
 and similar generalizations can be obtained for the stresses and moments,
\rf{th1}--\rf{ps2}.

 We cannot derive a governing equation,  similar to \rf{plst5}, for example, since now the 
in-plane 
 dependence of the stress and hence $w$ has been assumed {\it a~priori}.  However, the 
in-plane 
 wavenumber can be obtained in a self-consistent manner from the latter equation by assuming 
$w = w_0 e^{i\zeta x}$, where $w_0$ is constant.  This yields an equation for $\zeta$, 
\beq{zet}
\zeta^4 -  \omega^2 (1-\nu^2)\frac{\rho h}{EI} \left\{
 1 +  \left(\frac{1+\nu}{1-\nu}\right)\frac{E\theta_a \alpha^2}{C_p } \frac{k^2}{\gamma^2} 
f(\gamma h) \right\}^{-1} = 0, 
\eeq
the solution of which we  discuss next.

\subsection{Effects of transverse TE dissipation}

The wavenumber of a flexural wave in an undamped thin plate is $k_f$, where  
\beq{f1a}
k_f^4 =    \omega^2 (1-\nu^2){\rho h}/({EI}) , 
\eeq
 Treating $\zeta$ as an asymptotic series in the small parameter $\epsilon$, it is clear that 
 the leading order solution to the general dispersion relation of  \rf{zet} is $\zeta = 
 k_f +$O$(\epsilon)$.  The next term is given by 
\beq{zet2}
\zeta = k_f   \left[
1  -  \frac{\epsilon}{4} \left(\frac{1+\nu}{1-\nu}\right)\frac{k^2}{\gamma_f^2} f(\gamma_f h) 
+ \mbox{O}(\epsilon^2)\right], 
\eeq
where $\gamma_f = ( k^2 - k_f^2)^{1/2}$. 
Direct substitution gives  
\beq{gamf1}
\gamma_f = k_0 \left( 1 + i a l_{{\rm mfp}}/{h} - i\omega \tau_r\right)^{1/2},  
\eeq
where $k_0 = (-i\omega{C_p}/{K_3})^{1/2}$
and ${l_{{\rm mfp}}}$, the mean free path for phonons at temperature $T$, is
\beq{hstar} 
{l_{{\rm mfp}}}(T) = \frac{3K_3(T)}{\bar c \, C_p (T)}.
\eeq
The  quantity $\bar c$ is the average elastic wave speed, and the order one constant $a$ is 
given by 
\begin{equation}
\label{dp6}
a = \left[ 2(1-\nu )\rho /(3\mu )\right]^{1/2} \, (K_1/K_3)\, \bar c . 
\end{equation}
 Values of the mean free path  at room temperature are typically on the order of tens of 
nanometres using the value 
$\bar{c} = (c_L+2c_T)/3$ where $c_T=\sqrt{\mu/\rho}$ is the  transverse wave speed and $c_L = 
[2(1-\nu)/(1-2\nu)]^{1/2}\, c_T$ the longitudinal speed.    For isotropic conductivity the 
parameter  $a$ is a function of Poisson's ratio, 
\beq{para}
a = \left[\sqrt{8(1-\nu)}  + \frac{2(1-\nu)}{\sqrt{1-2\nu}}\right]/ \sqrt{27},
\eeq
which is of order unity.

 Equation \rf{gamf1} provides an avenue to compute corrections to classical TE dissipation 
 arising both from transverse diffusion and thermal relaxation.  
 The asymptotic solution \rf{zet2} indicates a propagating flexural wave has attenuation 
 equal to ${\rm Im}\,\zeta $.   Using the relation between attenuation and $Q$, and noting 
that the 
 undamped flexural wave has group velocity $2\omega/k_f$, we find 
 $1/Q = {\rm Im}\, 4{\zeta}/k_f$, or 
\beq{z3}
Q^{-1} = -  \epsilon  \left(\frac{1+\nu}{1-\nu}\right) \, {\rm Im} \, \bigg\{
{1 - i\omega \tau_r \over 1 - i\omega \tau_r +  ia {l_{{\rm mfp}}}/h} \, 
f \left( k_0 (1 - i\omega \tau_r+ia {l_{{\rm mfp}}} /h)^{1/2}\right) \bigg\} . 
\eeq
 In order to proceed further in understanding this we replace the function $f$ by its series 
 expansion. 
Equations   \rf{ap1}, \rf{ap3} and \rf{note} imply 
\beq{f11}
f\left( k_0 (1 - i\omega \tau_r+ia {l_{{\rm mfp}}}/h)^{1/2}\right)= 
\sum\limits_{n=0}^\infty  \, {f_n} \big[
 1 +  \frac{1} {-1+ i\omega \tau_n (1 - i\omega \tau_r+ia {l_{{\rm mfp}}}/h)} \big] \, . 
\eeq
where $\tau_n$, $n=0,1,2,\dots $ are defined in  $\rf{mod}_2$. 
Thus, 
\beq{z4}
Q^{-1} =   \epsilon  \left( \frac{1+\nu}{1-\nu}\right) \, \sum\limits_{n=0}^\infty  \, {f_n}\, 
\frac{\omega \tau_n\, \left( 1 + \omega \tau_n (a {l_{{\rm mfp}}}/h) \right) }
{\left( 1 + \omega \tau_n (a {l_{{\rm mfp}}}/h) - \omega^2 \tau_n \tau_r \right) ^2  +\omega^2 
\tau_n^2}. 
\eeq

The classical result for the TE loss in a vibrating beam is \cite{Zener}
\beq{z5}
Q^{-1} =   \epsilon   \, 
\frac{\omega \tau_0 }
{1 +\omega^2 \tau_0^2}. 
\eeq
 This ignores the effect of $\tau_r$, and for the purpose of comparison with \rf{z4} we 
 consider the case $\tau_r = 0$ in the latter.  We truncate \rf{z4} at the first term (which 
 is reasonable considering $f_0/f_1 = 81$), and for further simplicity, the  factor $f_0 = 
 {96}/{\pi^4} = 0.9855$ is replaced by unity, giving    
\beq{z6}
Q^{-1} =   \epsilon  \left(\frac{1+\nu}{1-\nu}\right) 
\frac{\omega \tau_0\, \big(1 + \omega \tau_0 (a {l_{{\rm mfp}}}/h)\big) }
{\big(1 + \omega \tau_0 (a {l_{{\rm mfp}}}/h)  \big)^2  +\omega^2 \tau_0^2}. 
\eeq
 Comparing  \rf{z5} and \rf{z6} the first difference is the factor $({1+\nu})/({1-\nu})$ 
 which can be attributed to the fact that it is a plate rather than a beam as discussed in 
 the previous subsection, see  \rf{mpf} and \rf{dp5}.  The major additional 
 distinction between the present theory and the classical result is the presence of the terms 
 involving ${l_{{\rm mfp}}}$ resulting from transverse diffusion, which are absent in Zener's theory 
 and previous analyses.  These corrections are always small, since the domain of validity of 
 the thermodynamic analysis employed here is restricted to distances far greater than the 
 mean free path, the domain $h \gg {l_{{\rm mfp}}}$.

 Alternatively, we note that \rf{z6} contains  two characteristic times:  the 
 `Zener' relaxation time $\tau_0 = h^2C_P/\pi^2 K_3$,  and a new characteristic time defined 
 by $\tau^* = \tau_0 (a {l_{{\rm mfp}}}/h)$.  It is easily checked that $\omega \tau^* = (k_f h/\pi 
 )^2$, which must be a small quantity in order for the Kirchhoff assumption to remain valid, 
 that is, a necessary prerequisite for the validity of the plate theory is that the 
flexural 
 wavelength is much longer than the thickness.  Also, 
\beq{ts}
\tau^* = {h}{\pi^{-2}} \sqrt{ 12(1-\nu^2)\rho/E}, 
\eeq
 which suggests interpreting $\tau^*$ as the travel time of an elastic wave across the 
 thickness.  But this time must be much less than $1/\omega$, otherwise thickness resonances 
can occur, again violating the conditions of the plate theory.

 Thus, the restrictions on the use of  both the thermal conduction  and the thin plate 
theories requires that 
$\omega \tau_0 (a {l_{{\rm mfp}}}/h) $ is small.  
 Nevertheless, small corrections arising from transverse diffusion can be estimated in this 
fashion.

 The effect of non-zero $\tau_r $ can be considered by using an estimate for this relaxation 
 time.  Rudgers \cite{rudgers90} provides the estimates  $\tau_r = \tau_{{\rm mfp}}/3$, where 
 $\tau_{{\rm mfp}} = {l_{{\rm mfp}}}/{\bar c}$ (Rudgers actually calculates two approximations for 
$\tau_r$ 
 but they are of the same order of magnitude).  The term $\omega \tau_r$ must remain small in 
 the context of thin plate theory, otherwise the same assumptions as before are violated, 
 for example, the wavelength is far less than the thickness. However, if the phonon mean 
free path is 
 comparable with the plate thickness, then both the transverse diffusion and the 
 Cattaneo--Vernotte terms can become important.

\section{Conclusion}

 Equation \rf{a13} is one of the basic results of the paper, as it provides a means to 
 compute TE dissipation given a solution in terms of the inhomogeneous stress.   This 
 computation is in general a complicated undertaking because it requires the determination of 
 the eigenfunctions and eigenvalues of the heat equation in the particular geometry of 
 interest.  Nevertheless, a clear recipe for an approximate calculation of the TE dissipation 
 in an arbitrary, anisotropic, elastic medium is given.

 The theoretical scope is then narrowed to the case of thin plate structures for which the 
 thermal heat flow simplifies dramatically.  In this case, we are able to obtain an explicit 
 formula for the TE loss in an arbitrary, anisotropic elastic system,~\rf{doug3}.  
 This equation, which shows that TE loss is in general inhomogeneous, is the other principal
 result obtained.
 It should be emphasized that the  $\Delta \cal E $ derived in   \rf{doug3} has the  property  
 that it is not a global measure of damping but is local, and can be used to define 
 TE loss at a point in a structure, and then integrated to determine the $Q$ of 
an 
 arbitrary vibrational mode.  This distinction is particularly important for geometries of 
 interest in MEMS/NEMS applications, for the TE loss rate of low order modes in 
 typical geometries may vary significantly with position.  The local result for TE loss 
 predicts that the attenuation of a flexural wave in a large plate is $(1+\nu)/(1-\nu)$ times 
 the attenuation of a wave in a beam at the same frequency.

 Finally, we have investigated the effects of transverse diffusion.  Transverse diffusion 
 effects are most easily analyzed for plane wave behaviour.  For this case  we have derived 
an 
 effective plate equation including the effects of thermoelasticity to leading order in the 
 TE coupling.  The TE dissipation can be examined fairly simply in the plane 
wave 
 case because TE losses are homogeneous, and are therefore contained in a single loss factor.  
 Corrections to TE loss resulting from transverse diffusion are found  to be generally small 
 within the domain of validity of our theoretical models.

\section*{Acknowledgement}

DMP gratefully acknowledges the support of the Office of Naval Research.

% \bibliography{thermoelastic}

\filbreak %%%%%%%%%%%%%%%%%%%%%%%%%%%%%%%%%%%%%%%%%%%%%

\appendix
\renewcommand{\theequation}{{\rm A}.\arabic{equation}}
  \setcounter{equation}{0}

\section{General solutions of the equations of thermoelasticity}
 
 The governing equations \rf{sin} and \rf{cs} are more commonly expressed as coupled 
 equations of temperature and  displacement.  This can be achieved by first eliminating 
 $\bosy{\sigma}$ explicitly: 
\beqa{aa1}
\mbox{div} {\bf C}{\bf e} - \rho \ddot{\bf u} - \mbox{\boldmath$\beta$} \nabla \theta &=& 0,
\\
 \dot{\theta} + \tau_r \ddot{\theta} - C_v^{-1} \mbox{div}\ {\bf K} \nabla \theta
+     \big( 1+ \tau_r\frac{\partial \, }{\partial t} \big) \frac{\theta_a}{C_v} 
\mbox{\boldmath$\beta$}\cdot  \dot{{\bf e}}&=&0. 
\eeqa
 These may be expressed more succinctly as follows for time harmonic motion ($e^{-i\omega t}$ 
 is assumed), 
\beq{aa4}
{\cal L} (\nabla,\, \omega) {\bf U} + \omega^2 {\bf U} = 0,
\eeq
where 
\beqa{aa5}
  {\bf U} &=& 
\left\{ \ba{c}
\rho^{1/2}\ {\bf u} \\  
(-i\omega)^{-1}\ (C_v/\theta_a )^{1/2}\ \theta 
\ea \right\}\, \\ 
&& \nonumber \\
 {\cal L}  &=& 
\left[\ba{cc} 
\rho^{-1} \ {\bf Q}(\nabla) & i\omega (\rho C_v/\theta_a)^{-1/2}\ {\bf b}(\nabla) \\
 & \\
  i\omega (\rho C_v/\theta_a)^{-1/2} \ {\bf b}^T(\nabla) & (\theta_a/C_v)\  \tilde{\kappa} 
(\nabla ,\, \omega)
\ea \right],  
\eeqa
$Q_{ij}({\bf v}) = C_{ikjl}\ v_k v_l$,
${\bf b}({\bf v}) = \bosy{\beta}{\bf v}$ and
$\tilde{\kappa}  ({\bf v}, \omega) =
[ \theta_a ( \tau_r+(-i\omega)^{-1})]^{-1}\ 
{\bf v}\cdot  {\bf K} {\bf v}$.   Equation~\rf{aa4} represents a generalized eigenvalue 
problem for the complex-valued modal 
 frequencies.

 Some simplification is possible for isotropic bodies, for which $\bosy{\alpha} = \alpha {\bf 
 I}$ and $\bosy{\beta} = 3 \kappa_T \alpha {\bf I}$ where $\kappa_T = E/(1-2\nu) $ is the 
 isothermal bulk modulus.  Assuming the {\it ansatz} \cite{Chadwick62}
\beqa{aa6}
{\bf U} =  
\left\{ \ba{c}
\nabla \psi \\ 
\lambda \psi  
\ea \right\}, \qquad \mbox{where}\quad  \nabla^2 \psi + \Lambda^2 \psi = 0, 
\eeqa
the eigenvalue problem \rf{aa4} becomes 
\beqa{aa7}
 \omega^2 - c_L^2\Lambda^2 + \frac{i\omega \beta\lambda}{(\rho C_v/\theta_a)^{1/2}} &=& 0, \\
-  \frac{i\omega \beta\Lambda^2}{(\rho C_v/\theta_a)^{1/2}} + 
 \left( \omega^2 - \frac{K\Lambda^2}{C_v( \tau_r+(-i\omega)^{-1})}\right)
\lambda &=& 0, 
\eeqa
where $c_L$ is the longitudinal wave speed.  
 Eliminating  the constant $\lambda $ gives an equation for $\omega$ in terms of the 
wavenumber $\Lambda$: 
\beq{aa8}
 \big( \omega^2 - c_L^2\Lambda^2 \big) \left(  \frac{K\Lambda^2}{C_v( 1-i\omega\tau_r)} 
  - i\omega \right)
+ i  \omega \Lambda^2\frac{\theta_a \beta^2}{\rho C_v} = 0. 
\eeq
 Thus, the problem of determining the eigenfrequencies is reduced to finding solutions to the 
 Helmholtz equation \rf{aa6}$_2$ in the domain of interest.  Chadwick \cite{Chadwick62} 
 discusses this further and provides formulas for the roots of \rf{aa8} for  $\tau_r = 0$, in 
 which case it reduces to a cubic in $\omega$.

 The separation of variables approach does not work for generally anisotropic bodies.  
 However, it is worth noting that solutions of the form \rf{aa6} are  valid as long as the 
 thermal properties are isotropic and the elasticity tensor $\bf C$ satisfies 
 $C_{ijkl}n_jn_kn_l = c_0 n^2 n_i$ for all $\bf n$, and some constant $c_0$.  The most 
 general form of anisotropy with this property has stiffness (using Voigt notation) 
\beqa{aa9}
[{\bf C}] = \left[\ba{cccccc} 
c_0  & c_0-2c_{66} & c_0-2c_{55} & -2c_{56} &0 & 0
\\ & & & & & \\
c_0-2c_{66}  & c_0 & c_0-2c_{44} &0 &  -2c_{46} & 0 
\\  & & & & & \\
 c_0-2c_{55} & c_0-2c_{44} & c_0 & 0 & 0 & -2c_{45} 
  \\  & & & & & \\
-2c_{56} & 0 & 0 & c_{44} & c_{45} & c_{46}
 \\  & & & & & \\
0 & -2c_{46} &0  &  c_{45}  & c_{55} & c_{56}
\\  & & & & & \\
0 &0 &  -2c_{45} &  c_{46} & c_{56} & c_{66}
\ea \right] . 
\eeqa

\renewcommand{\theequation}{{\rm B}.\arabic{equation}}
  \setcounter{equation}{0}  
  
\section{Exact results for the function $f$}

Some exact results are  presented for the function $f$ of  \rf{aa3}  occurring in the TE 
theory for beams and plates in flexure. We begin with the representation 
\beq{ap1}
f( \xi ) 
=  1 + \sum\limits_{j=0}^\infty  \, \frac{f_j}{\frac12  \Omega_j  - 1} ,
\eeq
where  $f_j$, $j=0,\, 1,\, 2,\, \ldots$,  are defined in  \rf{inn} and 
\beq{ap3}
 \Omega_j = {2 \xi^2}/[(2j+1)\pi]^2. 
\eeq
 The infinite series in  \rf{ap1} is a consequence of the fact that the left member is a 
 meromorphic function of $\xi$ with simple poles 
 at $\xi = \pm (2j+1)\pi $, and residues that are readily determined.  Note that 
\beq{note}
\sum\limits_{j=0}^\infty  f_j = 1.
\eeq
It follows from \rf{ap1} that for real-valued $\xi$, 
\beq{ap4}
\mbox{Im}\,f\left( (1+i)(1-ir)^{1/2} \xi \right) 
= - \sum\limits_{j=0}^\infty  f_j \, \frac{\Omega_j}{(1-r \Omega_j)^2 + \Omega_j^2} . 
\eeq

We also note the identity 
\beq{ap8}
f( (1+i)\xi) 
 =1-  \frac{6}{\xi^2} \left\{i +\frac{(1-i)}{\xi}\big( \frac{\sinh \xi - i\sin \xi}{\cosh 
 \xi+\cos\xi }\big)\right\},
\eeq
 which is useful in the case of $\tau_r=0 \, (r =0)$, as it provides a closed form expression 
 for the TE damping in that case via \cite{Lifshitz00} 
\beq{ap9}
\mbox{Im}\, f( (1+i)\xi) 
 =- \frac{6}{\xi^2} \left\{1 -\frac{1}{\xi}\big( \frac{\sinh \xi + \sin \xi}{\cosh 
\xi+\cos\xi }\big)\right\}.
\eeq

 The exact result for a beam in flexure was derived by Lifshitz and Roukes \cite{Lifshitz00}, 
 and corresponds to the use of the function $f$ of  \rf{aa3}.  The  equivalence of their 
 derivation and  Zener's prediction \cite{Zener} is confirmed by  \rf{ap4} (both Zener  and 
 Lifshitz and Roukes considered the case $r=0$).  Thus, while the closed form expression of 
 Lifshitz and Roukes is novel, it is simply a more concise expression of the infinite series 
 of Zener.  Both are based on the same implicit or explicit first order  approximation 
 arising from the small difference between the adiabatic and isothermal systems.

 In fact, the analysis of Lifshitz and Roukes \cite{Lifshitz00} is a special case of the more 
 general problem solved by Alblas \cite{Alblas81}, in which the lateral (third) dimension of 
 the beam is taken into account and the boundary value problem for the temperature is more 
 general.   However, Alblas considered the situation in which the boundary condition for the 
 temperature variation is zero, and as a result he obtained a different functional form of 
 the TE damping.

\renewcommand{\theequation}{{\rm C}.\arabic{equation}}
  \setcounter{equation}{0}  
\section{Thermoelasticity theory for a circular rod}
 
In the case of a circular rod of radius $a$ we find that the solution to \rf{sin2} is 
\beq{c1}
\theta = -   (I_c c_p)^{-1} \langle z,\, \bosy{\alpha}\cdot  \bosy{\sigma}\rangle\, 
\left( r - \frac{J_1( kr)}{k J_1'(ka)}  \right)\cos \psi , 
\eeq
 where $I_c = \langle z,\, z\rangle\ = \pi a^4/4$, $z=r\cos \psi$ and $k$ is defined in 
 \rf{sin3}.  The function associated with the moment of the temperature is $f(ka)$ where now  
\beq{c2}
 f(\zeta) = 1+ \frac{4}{\zeta^3}\bigg[ \zeta  -  \frac{J_1(\zeta)}{J_1'(\zeta)} \bigg]\, . 
\eeq
 Zener \cite{Zener38} analysed  the circular rod  using the projection method involving a 
 series rather a closed form expression.  By comparing the results here with those of Zener, 
 we conclude that  
\cite{Zener38}
\beq{38}
 f_n= \frac{8}{q_n^2(q_n^2 - 1)}, \quad \tau_n = \frac{a^2C_p}{q_n^2K}, \, \,  
\mbox{where}\quad J_1'(q_n)=0, \, \, 
n=0,1,2,\ldots \, . \quad
\eeq

\end{document}